\documentclass[12pt]{article}

\usepackage{graphics}
\usepackage{epsfig}
\usepackage{cite}
\input epsf

\title{Prompt $J/\psi$ production at LHC: new evidence \\ for the $k_T$-factorization}
\author{S.P.~Baranov$^a$, A.V.~Lipatov$^b$, N.P.~Zotov$^b$}

\begin{document}

\maketitle

\begin{center}

{\it $^a$\,P.N.~Lebedev Physics Institute,\\ 
119991 Moscow, Russia\/}\\[3mm]

{\it $^b$\,D.V.~Skobeltsyn Institute of Nuclear Physics,\\ 
M.V. Lomonosov Moscow State University,
\\119991 Moscow, Russia\/}\\[3mm]

\end{center}

\vspace{1cm}

\begin{center}

{\bf Abstract }

\end{center}

In the framework of the $k_T$-factorization approach, 
the production and polarization of prompt $J/\psi$ mesons in $pp$ collisions at the LHC 
energy $\sqrt s = 7$~TeV is studied.
Both the direct production mechanism as well as 
feed-down contributions from $\chi_{c1}$, $\chi_{c2}$ and 
$\psi^\prime$ decays are taken into account.
Our consideration is based on the color singlet model supplemented
with the off-shell matrix elements for the corresponding partonic subprocesses.
The unintegrated gluon densities in a proton are determined using
the CCFM evolution equation as well the Kimber-Martin-Ryskin prescription.
We compare our numerical predictions with the first
experimental data taken by the CMS, ATLAS and LHCb collaborations.
The estimation of polarization parameters $\lambda_\theta$, 
$\lambda_\phi$ and $\lambda_{\theta \phi}$ which determine $J/\psi$ spin density matrix is performed.

\vspace{1.0cm}

\noindent
PACS number(s): 12.38.-t, 13.20.Gd, 13.88.+e

\vspace{1.0cm}

\section{Introduction} \indent 

The production of charmonium states at high energies is under 
intense theoretical and experimental study\cite{1,2,3}. The production mechanism
involves the physics of both short and long distances,
and so, appeals to both perturbative and nonperturbative methods of QCD. 
This feature gives rise to two competing theoretical approaches known 
in the literature as the color singlet (CS)\cite{4} and color octet (CO)\cite{5} models.
In the CS model, only those states 
with the same quantum numbers as the resulting charmonium contribute to the 
formation of a bound state. This is achieved by radiating a hard gluon in a
perturbative process. 
In the CO model, it was suggested to 
add the contribution of transition mechanism from $c\bar c$ pairs to charmonium, 
where a charmed quark pair is produced in a color octet 
state and transforms into the final color singlet state by the help of 
soft gluon radiation. The CO model is based on the general principle of 
the non-relativistic QCD factorization (NRQCD)\cite{6}.
As it is well known, the sole leading order (LO) CS model 
is insufficient to describe the experimental data on the $J/\psi$ production at the Tevatron energies.
By adding the contribution from the octet states and fitting the free parameters 
one was able to describe the data on the $J/\psi$ production at energies of modern colliders (see \cite{7}
and references therein).

However, recently the next-to-leading order (NLO)\cite{8} and dominant next-to-next-to-leading order (NNLO$^*$)\cite{9}
corrections to the CS mechanism have been calculated and have been
found to be essential in description of quarkonia production.
The comparison with the first LHC measurements performed by the ATLAS, CMS and LHCb 
collaborations demonstrates\cite{10} that the NNLO$^*$ CS model correctly reproduces the 
transverse momentum distributions as well as the total cross section
of $J/\psi$ mesons at $\sqrt s = 7$~TeV. 

The effect of high-order QCD corrections is also manifest in the polarisation
predictions. While the charmonium produced inclusively or in association with a photon are
predicted to be transversely polarised at LO, it has been found that their polarisation at NLO
is increasingly longitudinal at high $p_T$\cite{9, 11}.
Opposite, the NRQCD predicts the strong transverse polarization of the final state quarkonia\cite{1}.
This is in disagreement with the polarisation measurement\cite{12}
performed by the CDF collaboration at the Tevatron, casting doubt on the earlier conclusion 
that the CO terms dominate $J/\psi$ production.

The results of studies\cite{8,9,10,11} support the predictions\cite{13,14,15,16,17,18,19,20,21,22} obtained 
in the framework of the $k_T$-factorization QCD approach\cite{23}, 
where investigations of heavy quarkonia production and polarization have own long story.
Shortly, it was demonstrated\cite{14,15,16,17,18,19,20,21} that the experimental data on quarkonia
production at HERA, RHIC and Tevatron can be well described within
the CS model alone. The values of CO contributions obtained by fitting
the Tevatron data appear to be substantially smaller
than the ones in the NRQCD formlalism\cite{14, 16, 24}.
Furthermore, the longitudinal polarization of produced $J/\psi$ 
mesons predicted by the $k_T$-factorization is an immediate consequence 
of initial gluon off-shellness\cite{14} which taken into account in the $k_T$-factorization 
approach\footnote{A detailed description and discussion 
of the $k_T$-factorization approach can be found, for example, in reviews\cite{27}.}.

In the present note 
we give the systematic analysis\footnote{See also\cite{22}.} of first experimentl data\cite{26,27,28} on the 
prompt $J/\psi$ production taken by the CMS, ATLAS and LHCb collaborations at the LHC
energy $\sqrt s = 7$~TeV.
Follow the guideline of previous studies\cite{19,20}, in our consideration we will apply 
the CS model supplemented with the $k_T$-factorization approach. 
Two sources of $J/\psi$ production are taken into 
account: direct $J/\psi$ production and feed-down $J/\psi$ from the decay of other heavier prompt 
charmonium states like $\chi_{c1}$, $\chi_{c2}$ or $\psi^\prime$,
that is in a full agreement with the experimental setup\cite{26,27,28}.
Specially we concentrate on the $J/\psi$ spin alignment and
estimate three polarization parameters $\lambda_\theta$, 
$\lambda_\phi$ and $\lambda_{\theta \phi}$ defining
the spin density matrix of produced $J/\psi$ mesons.
As it was mentioned above, studies of polarization observables are useful
in discriminating the CS and CO production mechanisms,

The outline of our paper is following. In Section~2 we 
recall shortly the basic formulas of the $k_T$-factorization approach with a brief 
review of calculation steps. In Section~3 we present the numerical results
of our calculations and a discussion. Section~4 contains our conclusions.

\section{Theoretical framework} \indent 

The production of prompt $J/\psi$ mesons in $pp$ collisions at the LHC can
proceed via either direct gluon-gluon fusion or the production of heavier $P$-wave states 
$\chi_{cJ}$ ($J = 0, 1, 2$) and $S$-wave state $\psi^\prime$, followed by their radiative
decays $\chi_{cJ} \to J/\psi + \gamma$ and $\psi^\prime \to J/\psi + X$. 
In the CS model, the direct mechanism corresponds to the partonic subprocess $g^* + g^* \to J/\psi + g$ which includes
the emission of an additional hard gluon in the final
state. The production of $P$-wave quarkonia is given by
$g^* + g^* \to \chi_{cJ}$\cite{21} and there is no emission of any additional
gluons. The feed-down contribution from $S$-wave state $\psi^\prime$
is described by the $g^* + g^* \to \psi^\prime + g$ subprocess.

The production amplitudes of all these subprocesses 
can be obtained from the one for an unspecified $c\bar c$ state 
by the application of appropriate projection operators $J(S, L)$
which guarantee the proper quantum numbers of the $c\bar c$ state 
under consideration. 
These operators for the different spin and orbital angular momentum states can be written as\cite{4}
$$
  J(^3S_1) = J(S = 1, L = 0) = \hat \epsilon(S_z)(\hat p_c + m_c)/m^{1/2}, \eqno(1)
$$
$$
  J(^3P_J) = J(S = 1, L = 1) = (\hat p_{\bar c} - m_c) \hat \epsilon(S_z)(\hat p_c + m_c)/m^{3/2}, \eqno(2)
$$

\noindent 
where $m$ is the mass of the specifically considered $c\bar c$ state, $p_c$ and $p_{\bar c}$
are the four-momenta of the charmed quark and anti-quark.
In accordance with the non-relativistic formalism of bound state formation, the charmed
quark mass $m_c$ is always set equal to $1/2$ of the quarkonium mass.
States with various projections of the spin momentum onto the $z$ axis are represented 
by the polarization vector $\epsilon(S_z)$.

The probability for the two quarks to form a meson depends on the bound state wave 
function $\Psi(q)$. In the non-relativistic approximation, 
the relative momentum $q$ of the quarks in the bound state is treated as a small quantity.
So, we represent the quark momenta as follows:
$$
  p_c = p/2 + q, \, p_{\bar c} = p/2 - q, \eqno(3)
$$

\noindent 
where $p$ is the four-momentum of the final state quarkonium.
Then, we multiply the relevant partonic amplitude ${\cal A}$ (depending on $q$) 
by $\Psi(q)$ and perform integration with respect 
to $q$. The integration is performed after expanding the integrand around $q = 0$:
$$
  {\cal A}(q) = {\cal A}|_{q = 0} + q^\alpha (\partial {\cal A} / \partial q^\alpha)|_{q = 0} + ..., \eqno(4)
$$

\noindent 
Since the expressions for ${\cal A}|_{q = 0}$ and 
$\partial {\cal A} / \partial q^\alpha|_{q = 0}$ are
no longer dependent on $q$, they may be factored outside the
integral sign. A term-by-term integration of this series then
yields\cite{29}
$$
  \int {d^3 q\over (2 \pi)^3} \, \Psi(q) = {1\over \sqrt{4\pi}} {\cal R}(x = 0), \eqno(5)
$$
$$
  \int {d^3 q\over (2 \pi)^3} \, q^\alpha \Psi(q) = - i \epsilon^\alpha(L_z){\sqrt 3\over \sqrt{4\pi}} {\cal R^\prime}(x = 0), \eqno(6)
$$

\noindent 
where ${\cal R}(x)$ is the radial wave function in the coordinate representation,
i.e. the Fourier transform of $\Psi(q)$.
The first term in (4) contributes only to $S$ waves, but vanishes for $P$ waves 
because ${\cal R}_P(0) = 0$. On the contrary, the second term contributes only to 
$P$ waves, but vanishes for $S$ waves because ${\cal R^\prime}_S(0) = 0$. 
States with various projections of the orbital angular momentum onto the $z$ 
axis are represented by the polarization vector $\epsilon(L_z)$.
The numerical values of the wave functions are either known from the leptonic decay widths 
(for $J/\psi$ and $\psi^\prime$ mesons) or can be taken
from potential models (for $\chi_{cJ}$ mesons).

In our numerical calculations, 
the polarization vectors $\epsilon(S_z)$ and $\epsilon(L_z)$ are defined as
explicit four-vectors. In the frame where the $z$ axis is oriented along the 
quarkonium momentum vector $p^\mu = (E,0,0,|{\mathbf p}|)$, these polarization 
vectors read
$$
  \epsilon^\mu(\pm 1) = (0, \pm 1, i, 0)/\sqrt 2, \, \epsilon^\mu(0) = (|{\mathbf p}|, 0, 0, E)/m. \eqno(7)
$$

\noindent 
The states with definite $S_z$ and $L_z$ are translated into states 
with definite total momentum $J$ and its projection $J_z$ using the Clebsch-Gordan coefficients:
$$
  \epsilon^{\mu \nu}(J, J_z) = \sum_{S_z, \, L_z} \langle 1, L_z; 1, S_z | J, J_z \rangle \, \epsilon^\mu(S_z) \, \epsilon^\nu(L_z). \eqno(8)
$$

\noindent 
Further evaluation of all partonic amplitudes under consideration (including 
subsequent leptonic and/or radiative decays, of course) is straightforward
and was done using the algebraic manipulation systems {\textsc FORM}\cite{30}.
We do not list here the obvious expressions because lack of space,
but only mention several technical points. First,
in according to the $k_T$-factorization prescription\cite{23},
the summation over the incoming off-shell gluon polarizations is 
carried with $\sum \epsilon^\mu \epsilon^{*\, \nu} = {\mathbf k}_T^{\mu} {\mathbf k}_T^{\nu}/{\mathbf k}_T^2$,
where ${\mathbf k}_T$ is the gluon transverse momentum orthogonal to the beam axis.
In the collinear limit, when $|{\mathbf k}_T| \to 0$, this expression converges to the 
ordinary $\sum \epsilon^\mu \epsilon^{*\, \nu} = - g^{\mu \nu}/2$ after averaging on the azimuthal angle.
In all other respects the evaluation follows the standard QCD Feynman rules. 
Second, the spin density matrix of final $J/\psi$ meson is determined by the
momenta $l_1$ and $l_2$ of the decay leptons and is taken in the 
form
$$
  \sum \epsilon^\mu \epsilon^{*\, \nu} = 3 \left( l_1^\mu l_2^\nu + l_1^\nu l_2^\mu - {m^2\over 2} g^{\mu \nu} \right)/m^2. \eqno(9)
$$

\noindent 
This expression is equivalent to the standard one 
$\sum \epsilon^\mu \epsilon^{*\, \nu} = - g^{\mu \nu} + p^\mu p^\nu/m^2$ 
but is better suited for studying the polarization observables 
because it gives access to the kinematic variables
describing the orientation of the decay plane.
Third, when considering the polarization properties of $J/\psi$ mesons 
originating from radiative decays of $P$-wave states, we rely upon the 
dominance of electric dipole $E1$ transitions\footnote{The same approach has been applied\cite{19} 
to study the $\Upsilon$ production and 
polarization at the Tevatron.}.
The corresponding invariant amplitudes can be written as\cite{31}
$$
  i {\cal A}(\chi_{c1} \to J/\psi + \gamma) = g_1 \, \epsilon^{\mu \nu \alpha \beta} k_\mu \epsilon_\nu^{(\chi_{c1})} \epsilon_\alpha^{(J/\psi)} \epsilon_\beta^{(\gamma)}, \eqno(10)
$$
$$
  i {\cal A}(\chi_{c2} \to J/\psi + \gamma) = g_2 \, p^\mu \epsilon^{\alpha \beta}_{(\chi_{c2})} \epsilon_\alpha^{(J/\psi)} \left[ k_\mu \epsilon_\beta^{(\gamma)} - k_\beta \epsilon_\mu^{(\gamma)} \right], \eqno(11)
$$

\noindent
where $\epsilon_\mu^{(\chi_{c1})}$, $\epsilon_\mu^{(J/\psi)}$ and $\epsilon_\mu^{(\gamma)}$
are the polarization vectors of a corresponding spin-one particles
and $\epsilon_{\mu \nu}^{(\chi_{c2})}$ is its counterpart for a spin-two $\chi_{c2}$ meson,
$p$ and $k$ are the four-momenta of the decaying quarkonium and the emitted photon,
$\epsilon^{\mu \nu \alpha \beta}$ is the fully antisymmetric Levita-Civita tensor.
The dominance of electric dipole transitions for the charmonium family is supported 
by the experimental data taken by the E835 Collaboration at the Tevatron\cite{32}.
Since the electromagnetic branching ratio for $\chi_{c0} \to J/\psi + \gamma$ decay
is more than an order of magnitude smaller than those for $\chi_{c1}$ and $\chi_{c2}$, we neglect its 
contribution to $J/\psi$ production. 
As the $\psi^\prime \to J/\psi + X$ decay matrix elements are unknown,
these events were generated according to the phase space.

The cross section of $J/\psi$ production 
at high energies in the $k_T$-factorization approach
is calculated as a convolution of the off-shell 
partonic cross section and the unintegrated gluon 
distributions in a proton. 
The contribution from the direct production mechanism
can be presented in the following form:
$$
  \displaystyle \sigma(p p \to J/\psi + X) = \int {1\over 16\pi (x_1 x_2 s)^2 } \, f_g(x_1,{\mathbf k}_{1T}^2,\mu^2) f_g(x_2,{\mathbf k}_{2T}^2,\mu^2) \times \atop
  \displaystyle  \times |\bar {\cal M}(g^* + g^* \to J/\psi + g)|^2 \, d{\mathbf p}_{T}^2 d{\mathbf k}_{1T}^2 d{\mathbf k}_{2T}^2 dy dy_g \, {d\phi_1 \over 2\pi} {d\phi_2 \over 2\pi}, \eqno (12)
$$

\noindent
where $f_g(x,{\mathbf k}_{T}^2,\mu^2)$ is the
unintegrated gluon density,
${\mathbf p}_T$ and $y$ are the transverse momentum and rapidity
of produced $J/\psi$ meson, $y_g$ is the rapidity of outgoing gluon 
and $s$ is the $pp$ center-of-mass energy.
The initial off-shell gluons have a fraction $x_1$ and $x_2$ 
of the parent protons longitudinal 
momenta, non-zero transverse momenta ${\mathbf k}_{1T}$ and 
${\mathbf k}_{2T}$ (${\mathbf k}_{1T}^2 = - k_{1T}^2 \neq 0$, 
${\mathbf k}_{2T}^2 = - k_{2T}^2 \neq 0$) and azimuthal angles
 $\phi_1$ and $\phi_2$. For the production of $\chi_{cJ}$ mesons via $2 \to 1$ 
subprocess above we have
$$
  \displaystyle \sigma(p p \to \chi_{cJ} + X) = \int {2\pi\over x_1 x_2 s \, T} \, f_g(x_1,{\mathbf k}_{1T}^2,\mu^2) f_g(x_2,{\mathbf k}_{2T}^2,\mu^2) \times \atop
  \displaystyle  \times |\bar {\cal M}(g^* + g^* \to \chi_{cJ})|^2 \, d{\mathbf k}_{1T}^2 d{\mathbf k}_{2T}^2 dy \, {d\phi_1 \over 2\pi} {d\phi_2 \over 2\pi}, \eqno (13)
$$

\noindent
where $T$ is the off-shell gluon flux factor. In the present analysis we set it to be equal to $T = 2 \hat s$,
where $\hat s$ is the energy of partonic subprocess.
In (12) and (13), $|\bar {\cal M}(g^* + g^* \to J/\psi + g)|^2$ and 
$|\bar {\cal M}(g^* + g^* \to \chi_{cJ})|^2$ are the corresponding off-shell matrix elements squared
and averaged over initial gluon polarizations and colors.
The production scheme of $\psi^\prime$ meson is identical to that of $J/\psi$, and only the 
numerical value of the wave function $|{\cal R}(0)|^2$ is different (see below).

In the numerical calculations we have tested 
a few different sets of unintegrated gluon distributions involved in~(12) and (13). First of them (CCFM set A0) 
has been obtained\cite{33} from the CCFM equation
where all input parameters have been fitted to describe the proton structure function $F_2(x, Q^2)$.
Equally good fit of the $F_2$ data was obtained using different values for the soft cut 
and a different value for the width of the intrinsic ${\mathbf k}_{T}$ distribution 
(CCFM set B0). 
Also we will use the unintegrated gluons taken in the 
Kimber-Martin-Ryskin (KMR) form\cite{34}. The KMR approach is a formalism to 
construct the unintegrated parton distributions from well-known conventional ones. 
For the input, we have used recent leading-order Martin-Stirling-Thorn-Watt (MSTW) set\cite{35}.

The multidimensional integrations in~(12) and~(13) have been performed
by the means of Monte Carlo technique, using the routine \textsc{vegas}\cite{36}.
The full C$++$ code is available from the author on 
request\footnote{lipatov@theory.sinp.msu.ru}.

\section{Numerical results} \indent

We now are in a position to present our numerical results. First we describe our
input and the kinematic conditions. After we fixed the unintegrated
gluon distributions, the cross sections (12) and (13) depend on
the renormalization and factorization scales $\mu_R$ and $\mu_F$. 
Numerically, we set 
$\mu_R^2 = m^2 + {\mathbf p}_{T}^2$ and
$\mu_F^2 = \hat s + {\mathbf Q}_T^2$, where ${\mathbf Q}_T$ is the 
transverse momentum of initial off-shell gluon pair. 
Note that the choice of $\mu_R$ is the standard one for studying of the 
$J/\psi$ production whereas the special choice of $\mu_F$ is connected with the CCFM evolution
(see\cite{33}). Following to\cite{37}, we set $m_{J/\psi} = 3.097$~GeV, $m_{\chi_{c1}} = 3.511$~GeV,
$m_{\chi_{c2}} = 3.556$~GeV, $m_{\psi^\prime} = 3.686$~GeV and use the LO formula 
for the coupling constant $\alpha_s(\mu^2)$ with $n_f = 4$ quark flavours
at $\Lambda_{\rm QCD} = 200$~MeV, such that $\alpha_s(M_Z^2) = 0.1232$. 
The charmonia wave functions at the origin
of coordinate space are taken to be equal to 
$|{\cal R}_{J/\psi}(0)|^2 = 0.0876$~GeV$^3$\cite{37},
$|{\cal R^\prime}_{\chi}(0)|^2 = 0.075$~GeV$^5$\cite{38},
$|{\cal R}_{\psi^\prime}(0)|^2 = 0.0391$~GeV$^3$\cite{37}.
According to\cite{37}, the following branching fractions are used:
$B(\chi_{c1} \to J/\psi + \gamma) = 0.356$,
$B(\chi_{c2} \to J/\psi + \gamma) = 0.202$,
$B(\psi^\prime \to J/\psi + X) = 0.561$ and
$B(J/\psi \to \mu^+ \mu^-) = 0.0593$.

The results of our calculations are presented in Figs.~1 --- 3 in 
comparison with the CMS, ATLAS and LHCb data\cite{26,27,28}. 
The solid, dashed and dash-dotted
curves correspond to the results obtained using the 
CCFM A0, B0 and KMR gluon densities, respectively.
Everywhere, we separately show the contribution from the 
direct production mechanism taken solely (dotted curves).
In this case we apply the CCFM A0 gluon density for illustration.
It is clear that sole direct production is not sufficient 
to describe the LHC data. However, we obtain a good overall agreement of our predictions 
and the data when summing up the direct and feed-down contributions.
The latter is important and production of $J/\psi$ mesons via radiative 
decays of $\chi_{cJ}$ and $\psi^\prime$ 
mesons even dominates over the direct contribution at large transverse momenta.
The reason can be seen in the fact that the production of
$\chi_{cJ}$ states refers to much lower values of the final state 
invariant mass and therefore effectively probes small $x$ region, where the gluon 
distributions are growing up. 
The dependence of our numerical results on the unintegrated PDFs is rather weak
and the CCFM and KMR predictions are practically coincide. The difference between them 
can be observed at small $p_T$ or at large rapidities probed at the LHCb measurements.

Computations\cite{7} performed in the framework of NRQCD, where CO contributions
are taken into account, can also explain at satisfactory level 
the shape and the absolute normalization of the measured $J/\psi$ cross-sections.
However, as it was mentioned above, they predict a substantial transverse 
component for the polarisation of $J/\psi$ mesons at large $p_T$ which is not 
supported by measurements. We find that in the framework of the $k_T$-factorization approach
no need for a CO contributions in the description of $J/\psi$
production at the LHC. From the other side, the account of high-order 
corrections to the CS cross sections calculated in the collinear QCD factorization 
also leads to the significant improvements in description of the data: 
the upper bound of the NNLO$^*$ CS predictions is very close\cite{11} to the measurements\cite{26,27,28}
and agree much better (compared to the LO CS results) with the $k_T$-factorization calculations 
which incorporates a large part of collinear high-order corrections at LO level.

Note that the calculated cross sections 
of feed-down contributions from the $P$-wave states 
are free from singularities at small transverse momenta.
This contrasts with the collinear QCD factorization predictions, which are either 
unphysical or even divergent.

\begin{table}
\begin{center}
\begin{tabular}{|l|c|c|c|c|c|c|}
\hline
   & & & & & & \\
  Source & $\lambda_\theta$ (HX) & $\lambda_\phi$ (HX) &  $\lambda_{\theta \phi}$ (HX) & $\lambda_\theta$ (CS) & $\lambda_\phi$ (CS) &  $\lambda_{\theta \phi}$ (CS)\\
   & & & & & & \\
\hline
   & & & & & & \\
   Direct & $-0.15$ & $-0.09$ & $0.01$ & $0.20$ & $-0.22$ & $-0.01$\\
   & & & & & &\\
   Feed-down & $0.19$ & $0.14$ & $0.00$ & $0.35$ & $0.09$ & $0.00$\\
   & & & & & &\\
\hline
   & & & & & &\\
   Total & $-0.07$  & $-0.03$ & $0.01$ & $0.24$ & $-0.14$ & $-0.01$\\
   & & & & & &\\
\hline
\end{tabular}
\end{center}
\caption{The polarization parameters of prompt $J/\psi$ mesons calculated
  in the kinematical region of CMS and ATLAS measurements\cite{26,27}. The CCFM A0 gluon density is used.}
\label{table1}
\end{table}

Now we turn to the the $J/\psi$ polarization.
In general, the spin density matrix of a vector
particle depends on three parameters $\lambda_{\theta}$, $\lambda_\phi$ and $\lambda_{\theta \phi}$ which can be
measured experimentally. So, the double differential
angular distribution of the $J/\psi \to \mu^+ \mu^-$ decay products reads\cite{39}
$$
  {d\sigma \over d\cos \theta^* d\phi^*} \sim 1 + \lambda_\theta \cos^2 \theta^* + 
    \lambda_\phi \sin^2 \theta^* \cos 2 \phi^* + \lambda_{\theta \phi} \sin 2 \theta^* \cos \phi^*, \eqno(14)
$$

\noindent
where $\theta^*$ and $\phi^*$ are the polar and azimuthal angles of the decay lepton measured
in the $J/\psi$ rest frame. Since the polarization 
parameters $\lambda_\theta$, $\lambda_\phi$ and $\lambda_{\theta \phi}$
(which greatly affects on the cross sections) are not determined yet at the LHC, 
the results of measurements performed by the CMS, ATLAS and LHCb collaborations have been presented in a different ways.
So, in the ATLAS analysis\cite{27} the unknown $J/\psi$ polarization
has been treated as an additional source of systematic uncertainties.
Contrary, the CMS and LHCb collaborations quote their measurements\cite{26,28} for
different polarization scenarios: unpolarized ($\lambda_\theta = 0$), 
full longitudinal polarization ($\lambda_\theta = - 1$) and full transverse
$J/\psi$ polarization ($\lambda_\theta = 1$) in the Collins-Soper or the helicity
frames\footnote{The experimental data points in Figs.~1 and 3 correspond to the 
unpolarized scenario.}. Below we estimate the 
polarization parameters $\lambda_\theta$, $\lambda_\phi$ and $\lambda_{\theta \phi}$
in a whole kinematical regions regarding the CMS, ATLAS and LHCb measurements.
Our evaluation is generally followed the experimental
procedure. We have collected the simulated
events in the specified bins of $J/\psi$ transverse momentum $p_T$ and rapidity $y$,
generated the decay lepton angular distributions according
to the production and decay matrix elements,
and then applied a three-parametric fit based on~(14).
The estimated values of polarization parameters $\lambda_\theta$, $\lambda_\phi$ and $\lambda_{\theta \phi}$ 
in the helicity (HX) and Collins-Soper (CS) frames are listed in Tables~1 and 2.
We find that these parameters are the same
in the kinematical regions covered by the CMS and ATLAS collaborations.
In order to study the production dynamics in more detail,
we separately show contributions from the direct and feed-down mechanisms.
The latter, of course, change the polarization of final $J/\psi$ mesons predicted by the
direct production mechanism\cite{19} but this effect is not well prononced 
due to overall integration over $J/\psi$ transverse momentum.
Note that the qualitative predictions
for the $J/\psi$ polarization are stable with respect to variations
in the model parameters. In fact, there is no dependence on
the strong coupling constant and unintegrated gluon densities, i.e. two of an important 
sources of theoretical uncertainties cancels out.
Therefore future precise measurements of the polarization parameters at the LHC 
will play crucial role in discriminating the different theoretical
approaches.

\begin{table}
\begin{center}
\begin{tabular}{|l|c|c|c|c|c|c|}
\hline
   & & & & & & \\
  Source & $\lambda_\theta$ (HX) & $\lambda_\phi$ (HX) &  $\lambda_{\theta \phi}$ (HX) & $\lambda_\theta$ (CS) & $\lambda_\phi$ (CS) &  $\lambda_{\theta \phi}$ (CS)\\
   & & & & & & \\
\hline
   & & & & & & \\
   Direct & $-0.03$ & $-0.13$ & $0.17$ & $0.19$ & $-0.22$ & $-0.03$\\
   & & & & & &\\
   Feed-down & $0.22$ & $0.11$ & $0.13$ & $0.43$ & $0.05$ & $0.05$\\
   & & & & & &\\
\hline
   & & & & & &\\
   Total & $0.03$ & $-0.07$ & $0.16$ & $0.26$ & $-0.14$ & $-0.01$\\
   & & & & & &\\
\hline
\end{tabular}
\end{center}
\caption{The polarization parameters of prompt $J/\psi$ mesons calculated
  in the kinematical region of LHCb measurements\cite{28}. The CCFM A0 gluon density is used.}
\label{table2}
\end{table}

\section{Conclusions} \indent 

We have investigated prompt $J/\psi$ production in $pp$ collisions
at the LHC energy $\sqrt s = 7$~TeV within the framework of the $k_T$-factorization 
approach. Both the direct production mechanism as well as 
feed-down contributions from $\chi_{c1}$, $\chi_{c2}$ and 
$\psi^\prime$ decays are taken into account.
Our consideration is based on the color singlet model supplemented with the 
off-shell matrix elements for the corresponding partonic subprocesses.
The unintegrated gluon densities in a proton are determined using
the CCFM evolution equation as well the Kimber-Martin-Ryskin prescription.
We have obtained well agreement of our calculations and the first 
experimental data taken by the CMS and ATLAS collaborations
when summing up the direct and feed-down contributions.
The dependence of our predictions on the unintegrated gluon
densities appears at small transverse momenta and at large rapidities covered by the LHCb experiment.
We have demonstrated also that in the framework of the $k_T$-factorization
there is no room for a color octet contributions for charmonium production at the LHC.

The estimation of the polarization parameters 
$\lambda_\theta$, $\lambda_\phi$ and $\lambda_{\theta \phi}$ which determine the 
$J/\psi$ spin density matrix is given.
The future experimental analysis of the quarkonium polarization at the LHC
turned out to be very important and informative for discriminating
the different theoretical models.

\section{Acknowledgements} \indent 
We are very grateful to 
DESY Directorate for the support in the 
framework of Moscow --- DESY project on Monte-Carlo
implementation for HERA --- LHC. 
A.V.L. was supported in part by the grant of the President of 
Russian Federation (MK-3977.2011.2).
Also this research was supported by the 
FASI of Russian Federation (grant NS-4142.2010.2), 
FASI state contract 02.740.11.0244 and 
RFBR grant 11-02-01454-a.

\newpage

\begin{figure}
\begin{center}
\epsfig{figure=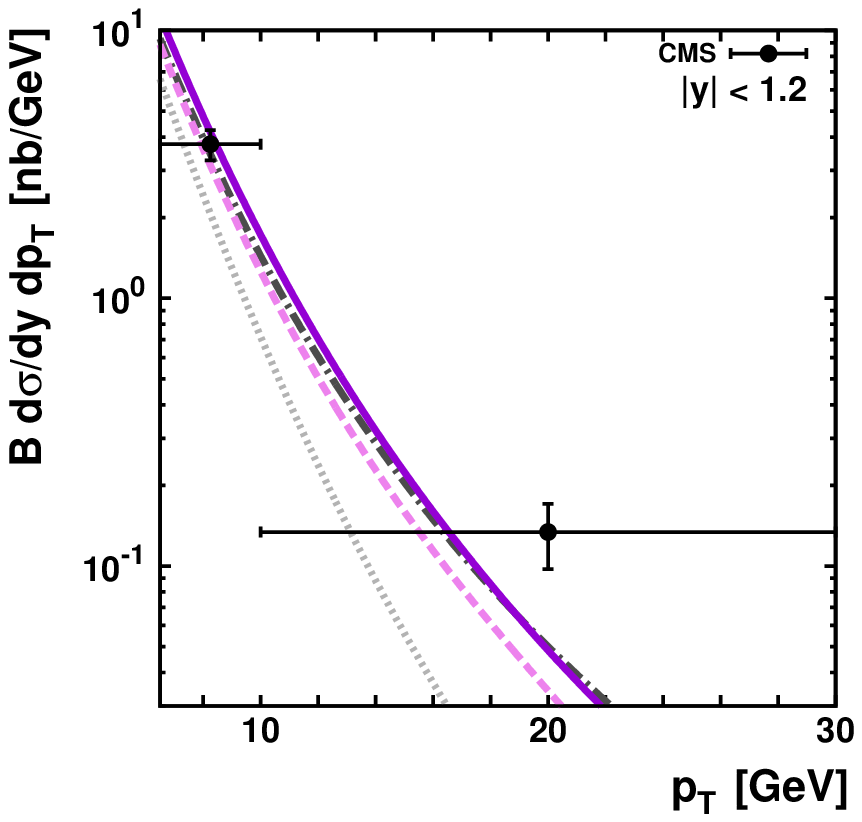, width = 8.1cm}
\epsfig{figure=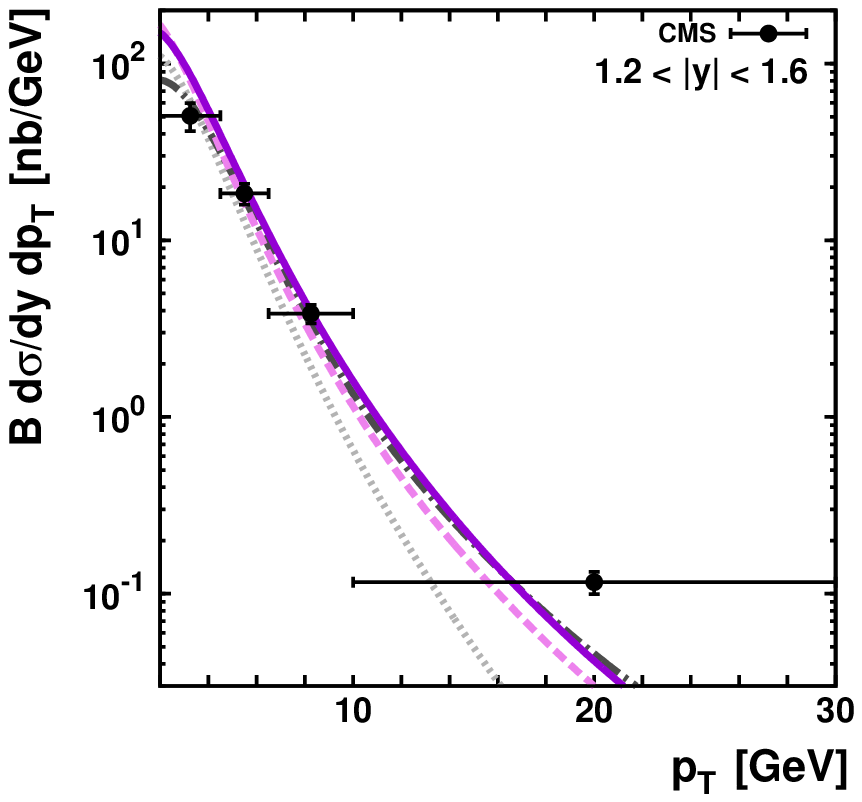, width = 8.1cm}
\epsfig{figure=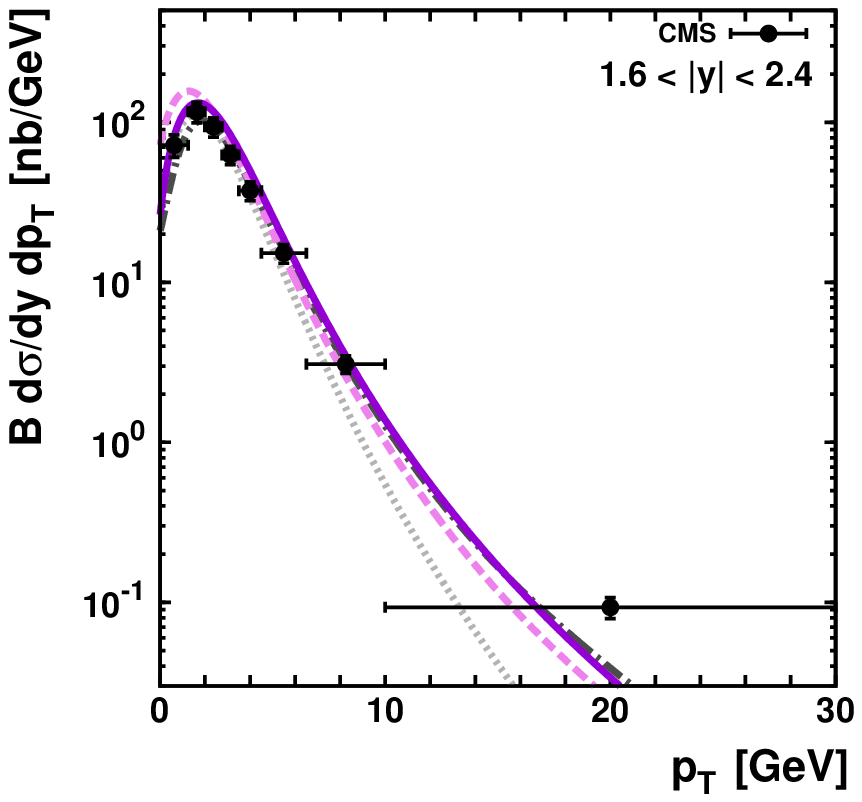, width = 8.1cm}
\caption{The double differential cross sections $d\sigma/dydp_T$ of prompt $J/\psi$ production
at $\sqrt s = 7$~TeV compared to the CMS data\cite{26}. The solid, dashed and dash-dotted
curves correspond to the results obtained using the CCFM A0, CCFM B0 and KMR gluon 
densities, respectively. The dotted curves represent the contribution from sole 
direct production mechanism calculated with the CCFM A0 gluon distribution.}
\end{center}
\label{fig1}
\end{figure}

\begin{figure}
\begin{center}
\epsfig{figure=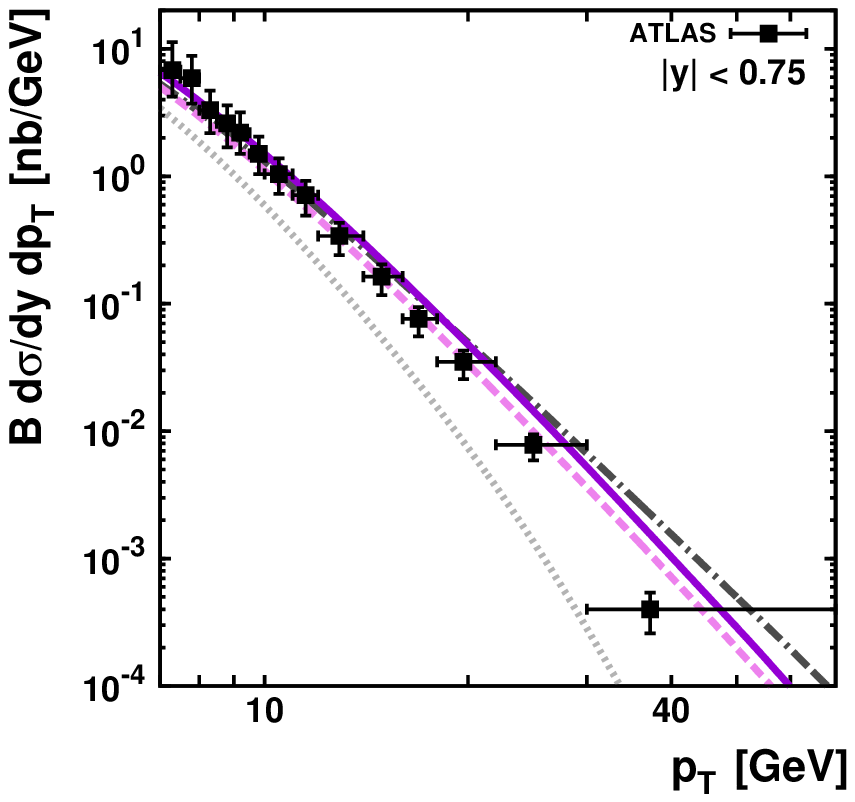, width = 8.1cm}
\epsfig{figure=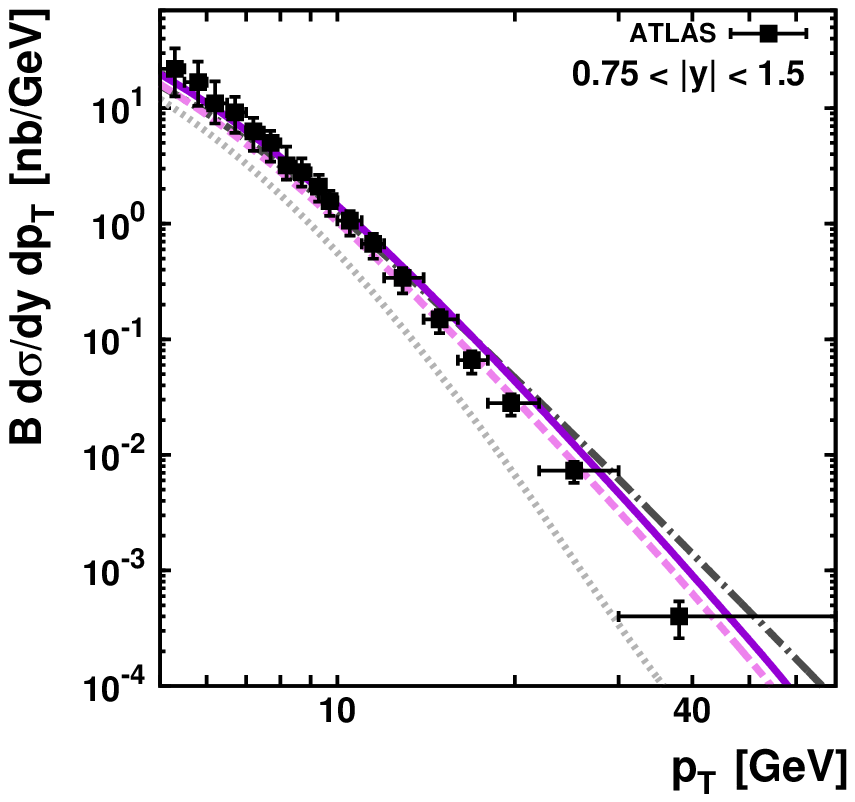, width = 8.1cm}
\epsfig{figure=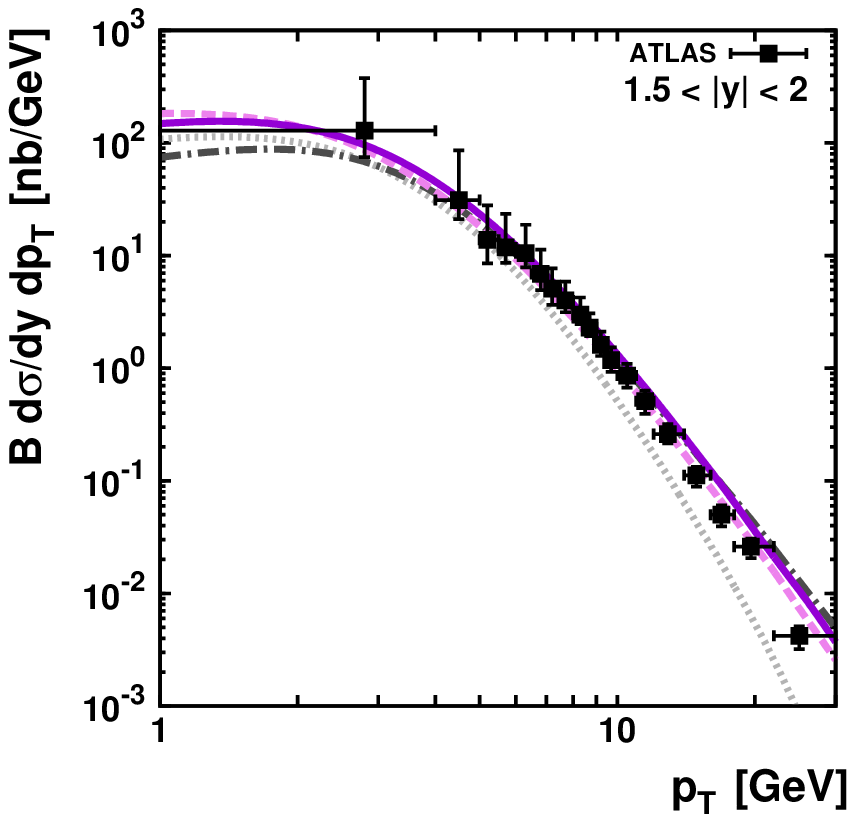, width = 8.1cm}
\epsfig{figure=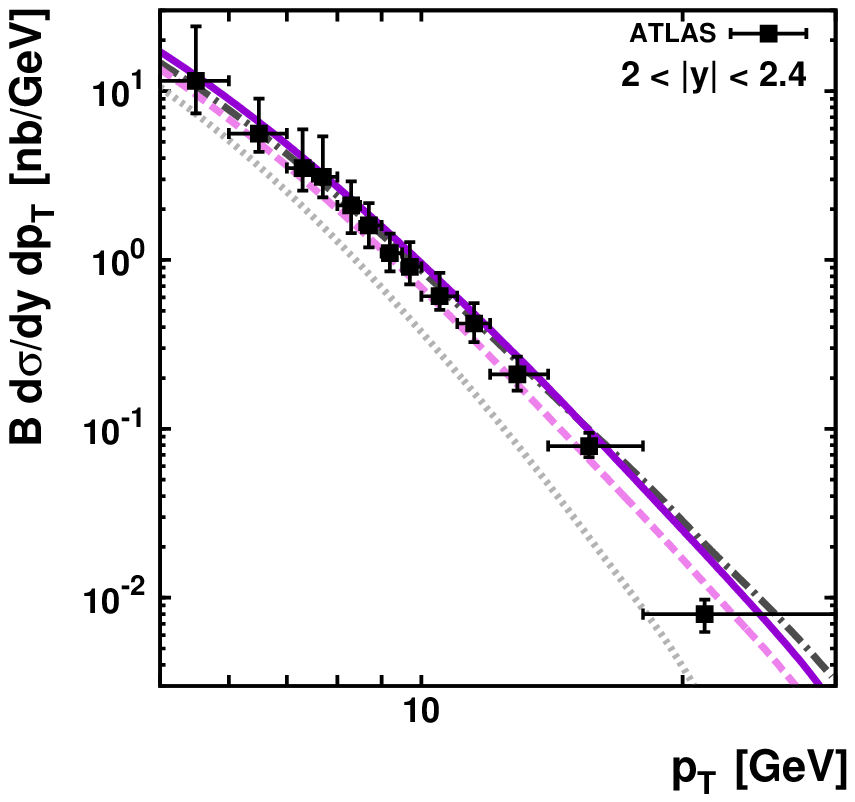, width = 8.1cm}
\caption{The double differential cross sections $d\sigma/dydp_T$ of prompt $J/\psi$ production
at $\sqrt s = 7$~TeV compared to the ATLAS data\cite{27}. 
Notation of all histograms is the same as in Fig.~1.}
\end{center}
\label{fig2}
\end{figure}

\begin{figure}
\begin{center}
\epsfig{figure=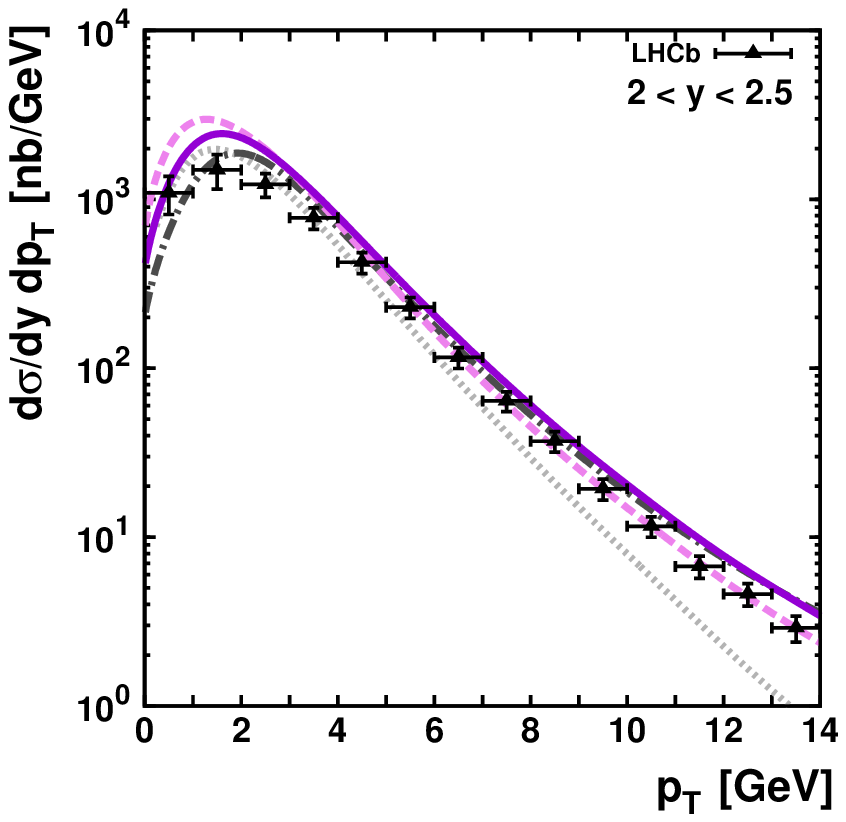, width = 8.1cm}
\epsfig{figure=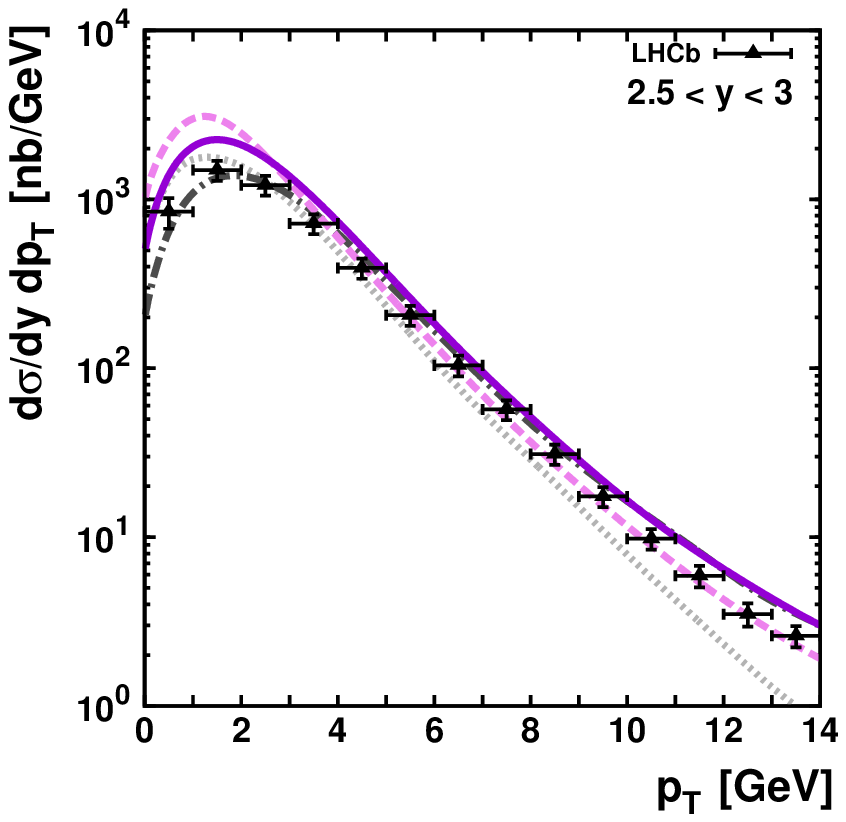, width = 8.1cm}
\epsfig{figure=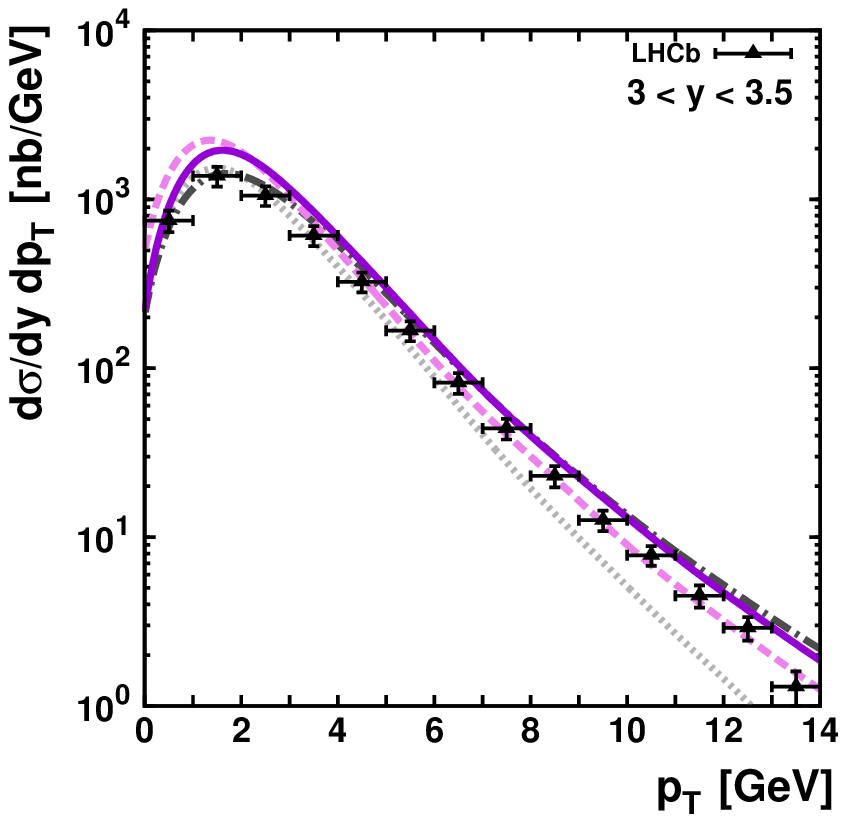, width = 8.1cm}
\epsfig{figure=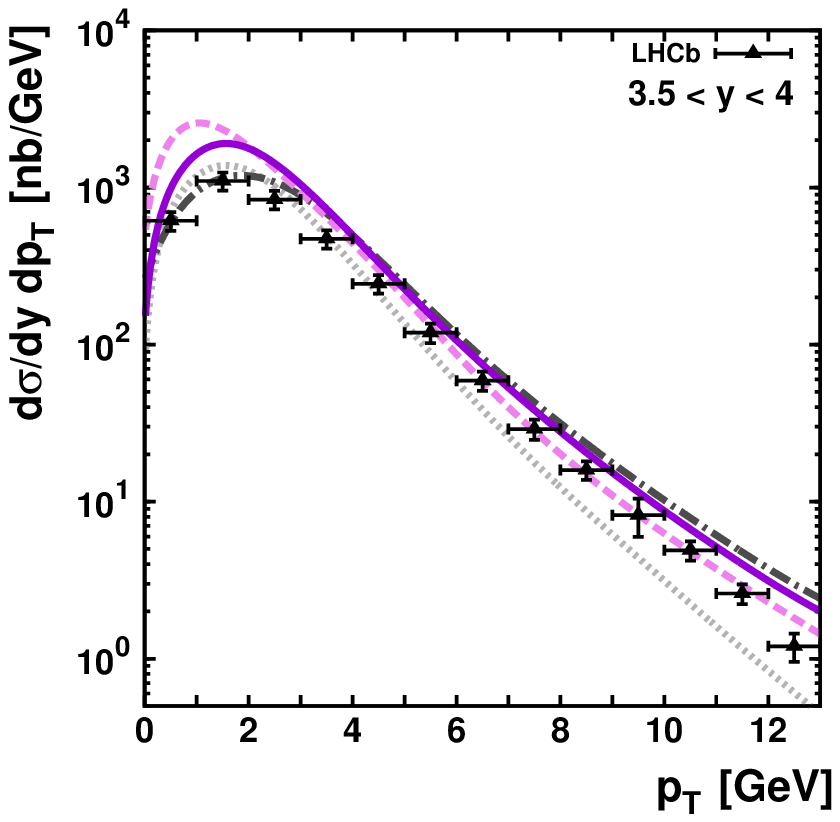, width = 8.1cm}
\epsfig{figure=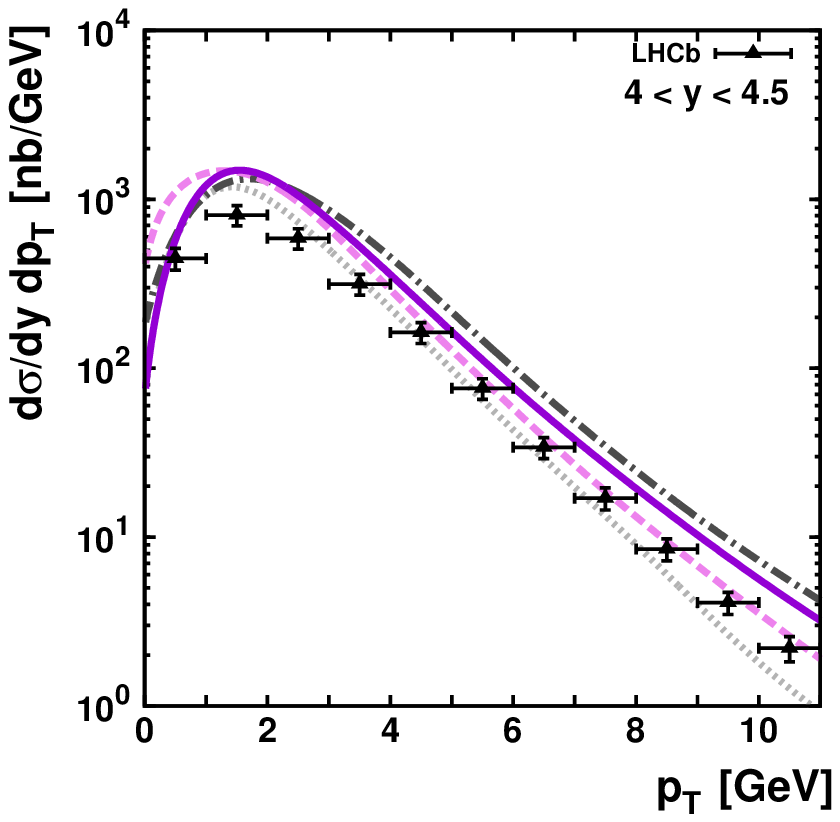, width = 8.1cm}
\caption{The double differential cross sections $d\sigma/dydp_T$ of prompt $J/\psi$ production
at $\sqrt s = 7$~TeV compared to the LHCb data\cite{28}. 
Notation of all histograms is the same as in Fig.~1.}
\end{center}
\label{fig3}
\end{figure}

\end{document}